\RequirePackage{fix-cm}
\documentclass[smallextended]{svjour3}       
\smartqed  
\usepackage{amsmath}
\usepackage{amssymb}
\usepackage{latexsym}
\usepackage{graphicx}
\begin{document}

\title{Quasi Normal Modes and P-V Criticallity for scalar perturbations in a class of dRGT massive gravity around Black Holes
}

\titlerunning{QNMs and P-V criticallity of Black Holes in dRGT Massive Gravity}        

\author{Prasia P         \and
        Kuriakose V C 
}


\institute{Prasia P. \at
              Department of Physics\\
              Cochin University of Science and Technology \\
              Cochin-682022\\
              \email{prasiapankunni@cusat.ac.in}           
           \and
           Kuriakose V. C. \at
              Department of Physics\\
              Cochin University of Science and Technology\\
              Cochin-682022\\
              \email{vck@cusat.ac.in}
}

\date{Received: date / Accepted: date}

\maketitle

\begin{abstract}
We investigate black holes in a class of dRGT massive gravity for
their quasi normal modes (QNMs) for neutral and charged ones using
Improved Asymptotic Iteration Method (Improved AIM) and their
thermodynamic behavior. The QNMs are studied for different values of
the massive parameter $m_g$ for both neutral and charged dRGT black
holes under a massless scalar perturbation. As $m_g$ increases, the
magnitude of the quasi normal frequencies are found to be
increasing. The results are also compared with the Schwarzchild de
Sitter (SdS) case. P-V criticallity of the aforesaid black hoels
under massles scalar perturbation in the de Sitter space are also
studied in this paper. It is found that the thermodynamic behavior
of a neutral black hole shows no physically feasible phase
transition while a charged black hole shows a definite phase
transition. 
\end{abstract}

\section{Introduction}
The  existence   of  black  holes   is  an  outcome  of  Einstein's
General Theory of  Relativity (GTR). The  question then  is how  to
realize their existence and one natural  way  to  identify  them is
to try to perturb  and know their  responses  to   the perturbation.
Regge and Wheeler\cite{1} started   way  back  in 1950s studying
perturbations  of black-hole  space  times and later, serious
studies were initiated by  Zerilli\cite{2}.  It was
Vishveshwara\cite{3} who first noticed the  existence  of
quasinormal modes (QNMs) by studying the scattering  of
gravitational  waves by Schwarzschild black holes. Later, scattering
of  scalar, electromagnetic  and Fermi fields  by different
black-hole spacetimes  have  been studied by
many\cite{ref7,ref8,ref9} and references cited therein. In the frame
work of general relativity, QNMs arise as perturbations of black
hole spacetimes. QNMs are the solutions to perturbation equations
and they  are distinguished from ordinary normal modes because they
decay at certain rates, having complex frequencies. The remarkable
property of the black hole QNMs("ring down" of black holes) is that
their frequencies are uniquely determined by the mass, angular
momentum and charge(if any) of black holes. Black holes can be
detected by observing  the QNMs through gravitational waves. When a
star collapses to form a black hole or when two black holes collide
or a black hole and a star collide, Gravitational Waves (GWs) are
emitted. The result of these processes is a black hole with higher
mass that absorbs the GWs \cite{6}. Hence the emitted GWs decay
quickly. The decay of oscillations are characterized by complex
frequencies.\\ \\
The Quasi normal modes were first introduced by Vishveshwara
\cite{7,8}. Later, perturbation calculations have been done by many
to get QNM oscillations \cite{9,10,11}. To study the black hole
QNMs, the solution of the perturbed field equation are separated for
the radial and angular parts, whose radial part is the so called
Regge-Wheeler equation. But this technique is time consuming and
complicated that makes it difficult to survey QNMs for a wide range
of parameter values. A semi analytic method has then been explored
\cite{12} that has its own limitations of accuracy. Later, the
Continued Fraction Method (CFM) was proposed by Leaver. This method
is a hybrid of analytic and numerical and can calculate QNM
frequencies by making use of analytic infinite series representation
of solution \cite{13}. Another method is WKB approximation which is
very commonly employed and a powerful one too. However all these
methods have their own limitations. In recent years a new approach
has been introduced to study black hole QNMs called Asymptotic
Iteration Method (AIM) which is previously used to solve eigenvalue
problems \cite{14}. This method has been shown to be efficient and
accurate for calculating
QNMs of black holes \cite{15,16}.\\ \\
The  studies of Hawking  and  Bekenstein  made  in 1970s\cite{17,18}
helped us to view that  black  holes  are  thermal objects
possessing temperature and  entropy   and  that laws  of black hole
dynamics are  analogous  to  the   laws  of classical
thermodynamics.  An immediate  consequence  of  these studies  is
that  they  bring together  quantum  theory,  gravity and
thermodynamics  and  one can  hope  for  a  quantum  theory of
quantum  gravity.  Various methods \cite{19,20} have  been developed
to study the  thermodynamics of  black  holes.  An important fact is
that certain  black  holes make a transition  from a stable phase to
an unstable  phase   and some  are thermodynamically
unstable\cite{21}. If the thermodynamic variables, pressure and
volume, are identified, then an equation of state corresponding to
the black hole can be found out and the critical points can be
determined. The P-V isotherms then show their thermodynamic
behavior. \\ \\
GTR   helped   us  to have a model for our universe and the universe
can  be considered as a dynamical system  and  most   of the
cosmological and astronomical observations could find a meaningful
explanations under GTR. But there are  some fundamental issues like
quantization of gravity, the initial  stages of  the evolution of
the Universe under Big-Bang   theory  and   also certain
astronomical observations like  dark matter  and  the late  time
accelerating expansion of the universe which  lacked  proper
explanations under GTR \cite{22,23}. Hence attempts   are  being
made  for  an alternative theory of gravitation.\\ \\
From the perspective of the modern  particle physics, GTR can be
thought of as the unique theory of a massless spin 2 particle called
graviton \cite{26,27,28}. If the assumption behind the uniqueness
theorem is broken, it can lead to alternative theories of gravity.
Theories concerning the breaking of Lorentz invariance and spin have
been explored in depth. Representing gravity as a manifestation of a
higher order spin, thereby maintaining the Lorentz invariance and
spin has also been explored largely \cite{29}. Yet another
possibility that has been recently explored is the so called
'Massive Gravity'(MG) theory\cite{30,ref12,ref13}. In this model
gravity is considered  to be propagated by a massive spin 2
particle. The theory gets complicated especially when the massive
spin 2 field interacts with matter. In that case, the theory goes
completely non-linear and consequently non renormalizable. A non
self interacting massive graviton model was first suggested by Fierz
and Pauli\cite{31}  which is now called as 'linear massive gravity'.
However this model suffers from a pathology\cite{32} thereby ruling
out the theory on the basis of solar system tests. Later,
Vainshtein\cite{33} proposed that the linear massive gravity model
can be recovered to GTR through 'Vainshtein Mechanism' at small
scales by including non linear terms in the hypothetical massive
gravity theory. But the Vainshtein mechanism is later found to
suffer from the so called 'Boulware-Deser'(BD) ghost\cite{34}.
Recently it is shown by de Rham, Gabadadze and Tolly in their series
of works \cite{35,36,37} that the BD ghost can be avoided for a sub
class of massive potentials. This is called   dRGT massive gravity
which includes one dynamical and one fixed metric. This
also holds true for its bi gravity extension \cite{30,32,ref3}.\\ \\
This paper deals with the study of quasinormal modes coming out of
massless scalar perturbations of a class of dRGT massive gravity
around both neutral and charged black holes. We use the Improved
Asymptotic Iteration Method (AIM) to calculate the QNMs. The P-V
criticality condition of such black holes are also verified in the
de Sitter space. Section 2 deals with a review of the Asymptotic
Iteration Method. In section 3, the quasinormal modes of neutral and
charged black holes coming under a class of dRGT massive gravity,
proposed by Ghosh, Tannukij and Wongjun \cite{39}, are found out.
Section 4 deals with the P-V criticality in the extended phase space
of black holes described in Section 3. Section 5 concludes the
paper.
%
%
\section{Review of Asymptotic Iteration Method}
Asymptotic Iteration Method(AIM) was proposed initially for finding
solutions of the second order differential equations of the form \cite{40},\\
\begin{equation}\label{}
    Y''(x)-\lambda_0(x)Y'(x)-s_0(x)Y(x)=0,\\
\end{equation}
where $\lambda_0(x)$ and $s_0(x)$ are coefficients of the
differential equation and are well defined functions and
sufficiently differentiable. By differentiating (1) with respect to
$x$,
\begin{equation}\label{}
    Y'''(x)-\lambda_1(x)Y'(x)-s_1(x)Y(x)=0,\\
\end{equation}
where the new coefficients are
$\lambda_1(x)=\lambda_0'+\lambda_0^2+s_0$ and
$s_1(x)=s_0'+s_0\lambda_0$. Differentiating $(1)$ twice with respect
to $x$ leads to,
\begin{equation}\label{}
    Y''''(x)-\lambda_2(x)Y'(x)-s_2(x)Y(x)=0,\\
\end{equation}
where the new coefficients are
$\lambda_2(x)=\lambda_1'+\lambda_1\lambda_0+s_1$ and
$s_2(x)=s_1'+s_0\lambda_1$. This process is continued to get the
$n^{th}$ derivative of $(1)$ with respect to $x$ as,
\begin{equation}\label{}
    Y^{(n)}(x)-\lambda_{n-2}(x)Y'(x)-s_{n-2}(x)Y(x)=0,\\
\end{equation}
where the new coefficients are related to the older ones through the
following expressions,
\begin{eqnarray}\label{}
    \lambda_n(x)=\lambda_{n-1}'+\lambda_{n-1}\lambda_0+s_{n-1},\\
    s_n(x)=s_{n-1}'+s_0\lambda_{n-1},
\end{eqnarray}
where $n=1,2,3,...$\\ \\
The ratio of $(n+2)^{th}$ derivative and $(n+1)^{th}$ derivative can
be obtained from $(4)$ as,
\begin{eqnarray*}
    \frac{Y^{(n+2)}(x)}{Y^{(n+1)}(x)}=\frac{d}{dx}(ln Y^{n+1})\\
    =\frac{\lambda_n [Y'(x)+\frac{s_n}{\lambda_n}Y(x)]}{\lambda_{n-1}[Y'(x)+\frac{s_{n-1}}{\lambda_{n-1}}Y(x)]}
\end{eqnarray*}
By introducing the asymptotic concept that for sufficiently large
values of $n$,
\begin{equation}
    \frac{s_n}{\lambda_n}=\frac{s_{n-1}}{\lambda_{n-1}}\equiv
    \alpha,
\end{equation}
where $\alpha$ is a constant, we get,\\
\begin{equation*}
    \frac{d}{dx}(ln Y^{n+1})=\frac{\lambda_n}{\lambda_{n-1}},
\end{equation*}
from which a general expression for $Y(x)$ can be found
out\cite{ref10}. From $(7)$ we can write,
\begin{equation}\label{}
    \lambda_n(x)s_{n-1}(x)-\lambda_{n-1}(x)s_n(x)=0.\\
\end{equation}
The roots of this equation are used to obtain the eigenvalues of
$(1)$. The energy eigenvalues will be contained in the coefficients.
To get the eigenvalues, each derivative of $\lambda$ and $s$ are
found out and expressed in terms of the previous iteration. Then by
applying the quantization condition given by $(8)$, a general
expression for the eigenvalue can be arrived at.  Cifti et al.
\cite{41} first noted that this procedure has a difficulty in that,
the process of taking the derivative of $s$ and $\lambda$ terms of
the previous iteration at each step can consume time and also affect
the numerical precision of calculations. To overcome this
difficulty, an improved version of AIM has been proposed that
bypasses the need to take derivative at each iteration. This is
shown to improve both accuracy and speed of the method. For that,
$\lambda_n$ and $s_n$ are expanded in a Taylor series around the
point at which AIM is performed, $x'$ ,\\
\begin{eqnarray}\label{}
    \lambda_n(x')=\sum_{i=0}^{\infty}c_n^i(x-x')^i,\\
    s_n(x')=\sum_{i=0}^{\infty}d_n^i(x-x')^i,
\end{eqnarray}
where $c_n^i$ and $d_n^i$ are the $i^{th}$ Taylor coefficients of
$\lambda_n(x')$ and $s_n(x')$ respectively. Substitution of
equations $(9)$ and $(10)$ in $(5)$ and $(6)$ lead to the recursion
relation for the coefficients as,
\begin{eqnarray}
  c_n^i &=& (i+1)c_{n-1}^{i+1}+d_{n-1}^i+\sum_{k=0}^ic_0^k c_{n-1}^{i-k}, \\
  d_n^i &=& (i+1)d_{n-1}^{i+1}+\sum_{k=0}^id_0^k c_{n-1}^{i-k},
\end{eqnarray}
Applying $(11)$ and $(12)$ in $(8)$, the quantization condition can
be rewritten as,
\begin{equation}\label{}
    d_n^0 c_{n-1}^0-d_{n-1}^0 c_n^0=0.
\end{equation}
This gives a set of recursion relations that do not require any
derivatives. The coefficients given by $(11)$ and $(12)$ can be
computed by starting at $n=0$ and iterating up to $(n+1)$ until the
desired number of recursions are reached. The quantization condition
given by $(13)$ contains only $i=0$ term. So, only the coefficients
with $i< N-n$ where $N$ is the maximum number of iterations to be
performed needs to be determined. The perturbed radial wave equation
of a black hole can be written in the form of a second order
differential equation similar to $(1)$ with the coefficients
containing their quasinormal frequencies. Hence the condition $(13)$
can be employed to extract the QNMs of a black hole\cite{15,16}.
This method is used in this paper to determine the QNMs of dRGT
black hole.
\section{Quasinormal modes of Black Holes in dRGT massive gravity}
\subsection{Neutral dRGT black hole}
In the standard formalism of dRGT massive gravity theory, the
Einstein-Hilbert action is given by \cite{42,ref14},
\begin{equation}\label{1}
    S=\int{d^{4}x}\sqrt{-g}\frac{1}{2\kappa^2}\left[R+m_{g}^{2}U(g,\phi)\right],\\
\end{equation}
where $g$ is the metric tensor, $R$ is the Ricci scalar, $m_g$
represents the graviton mass and $U$ is the effective potential for
the graviton and is given by \cite{43},
\begin{equation}\label{}
    U(g,\phi)=U_2+\alpha_3 U_3+\alpha_4 U_4,
\end{equation}
where $\alpha_3$ and $\alpha_4$ are two free parameters. These
parameters are redefined by introducing two new parameters $\alpha$
and $\beta$ as,
\begin{eqnarray}
  \alpha_3 &=& \frac{\alpha-1}{3}, \\
  \alpha_4 &=& \frac{\beta}{4}+\frac{1-\alpha}{12}.
\end{eqnarray}
Varying the action given by $(14)$ with respect to the metric leads
to the field equation,
\begin{equation}\label{}
    G_{\mu\nu}=\,-m^2 X_{\mu\nu},
\end{equation}
where,
\begin{equation}\label{}
    X_{\mu\nu}=\frac{\delta U}{\delta g_{\mu\nu}}-\frac{1}{2}U
    g^{\mu\nu}.
\end{equation}
The constraints of this field equation $(16)$ can be obtained by
using the Bianchi identity,
\begin{equation}\label{}
    \nabla^{\mu\nu}X_{\mu\nu}=0.
\end{equation}
A spherically symmetric metric  has a form given by,
\begin{equation}\label{}
    ds^2=g_{tt}(r)dt^2+2g_{tr}(r)dtdr+g_{rr}(r)dr^2+h(r)^2
    d\Omega^2,
\end{equation}
with $g_{tt}(r)=-\eta(r),$ $g_{rr}=\frac{1}{f(r)}$ and $h(r)=h_0 r$
where $h_0$ is a constant in terms of $\alpha$ and $\beta$
\cite{ref4,ref5,ref6}. The exact solution for this ansatz is
complicated. It is simplified by choosing specific relations for the
parameters. In this paper, we take $\alpha=-3\beta$. Since the
fiducial metric acts like a Lagrangian multiplier to eliminate the
BD ghost, to simplify the calculations, we choose the fiducial
metric as, \cite{44},
\begin{equation}\label{}
    f_{\mu\nu}=(0,0,c^2,c^2 \sin^2{\theta}),
\end{equation}
where $c$ is a constant.
In this paper we consider only the diagonal branch of the physical
metric for simplicity ie., $g_{tr}=0$. Then,
\begin{equation*}
    ds^2=-\eta(r)dt^2+\frac{dr^2}{f(r)}+r^2 d\Omega^2
\end{equation*}
By taking $\eta(r)=f(r)$ we get,
\begin{equation*}
    ds^2=-f(r)dt^2+\frac{dr^2}{f(r)}+r^2 d\Omega^2
\end{equation*}
The non-zero components of the Einstein tensor are given
by\cite{39},
\begin{eqnarray}
  G_t^t &=& \frac{f'}{r}+\frac{f}{r^2}-\frac{1}{r^2}, \\
  G_r^r &=& \frac{(rf'+f)}{r^2}-\frac{1}{r^2}, \\
  G^{\theta}_{\theta} &=& G^\phi_\phi ,\\
  &=&f'\left(\frac{f'}{4f}+\frac{1}{2r}\right)+\left(\frac{f''}{2}+\frac{f'}{2r}-\frac{(f')^2}{4f}\right),
\end{eqnarray}
and the $X_{\mu\nu}$ tensor as,
\begin{eqnarray}
  X^t_t &=& \left(\frac{\alpha(3r-c)(r-c)}{r^2}+\frac{3\beta(r-c)^2}{r^2}+\frac{3r-2c}{r}\right), \\
  X_r^r &=& -\left(\frac{\alpha(3r-c)(r-c)}{r^2}+\frac{3\beta(r-c)^2}{r^2}+\frac{3r-2c}{r}\right), \\
  X_\theta^\theta &=& X_\phi^\phi ,\\
   &=& \frac{\alpha(2c-3r)}{r}+\frac{3\beta(c-r)}{r}+\frac{c-3r}{r},
\end{eqnarray}
Solving $(18)$ using these expressions for $G_\mu^\nu$ and
$X_\mu^\nu$  gives the form of the metric as,
\begin{equation}\label{7}
    f(r)=1-\frac{2 M}{r}+\frac{\Lambda}{3}r^2+\gamma r+\zeta,
\end{equation}
where,
\begin{eqnarray}
  \Lambda &=& 3m_{g}^{2}\left(1+\alpha+\beta\right), \\
  \gamma &=& -cm_{g}^{2}\left(1+2\alpha+3\beta\right), \\
  \zeta &=& c^2m_{g}^2\left(\alpha+3\beta\right).
\end{eqnarray}
The details  of  the  above calculations are given by Ghosh,
Tannukij and Wagjun \cite{39}. When $\gamma=\zeta=0$, $\alpha$ and
$\beta$ will determine the nature of the solution. ie., if
$(1+\alpha+\beta)< 0$ we get a Schwarzschild-de Sitter type
solution, if $(1+\alpha+\beta)>0,$ we will get a Schwarzschild-anti
de Sitter type solution and when $m_g\rightarrow
0$ we get a Schwarzchild black hole.\\ \\
In this paper, we consider a static spherically symmetric space time
with vanishing Energy momentum tensor and hence the field
perturbations in such background are not coupled to the
perturbations of the metric and therefore are equivalent to test
field in black hole background. Consider a massless scalar field
that satisfies the Klein-Gordon equation in curved space-time,
\begin{eqnarray}\label{8}
    \frac{1}{\sqrt{-g}}\frac{\partial}{\partial x^{a}}g^{ab}\sqrt{-g}\frac{\partial}{\partial
    x^{b}}\Phi=0\\
    ie.,\frac{1}{f(r)}\frac{\partial^{2}\Phi}{\partial t^2}-\frac{\partial}{\partial
    r}f(r)\frac{\partial \Phi}{\partial
    r}-\frac{\Delta_{\theta,\phi}\Phi}{r^{2}}=0,
\end{eqnarray}
where,
\begin{equation}
\Delta_{\theta,\phi}=\frac{1}{\sin{\theta}}\frac{\partial}{\partial\theta}(\sin{\theta})+%
\frac{1}{\sin^{2}{\theta}}\frac{\partial^2}{\partial\phi^2}\\
\end{equation}
In order to separate out the angular variables we choose the ansatz:
\begin{equation}
    \Phi=\sum_{l=0}^{\infty}\sum_{m=0}^{l}\frac{R(r)}{r}e^{-i\omega
    t}Y_{l,m}(\theta,\phi),
\end{equation}
where $\omega$ gives the frequency of the oscillations corresponding
to the black hole perturbation, $Y_{l,m}(\theta,\phi)$ are the
spherical harmonics and,
\begin{equation}
    \Delta_{\theta,\phi}Y_{l,m}(\theta,\phi)=-l(l+1)Y_{l,m}(\theta,\phi).
\end{equation}
Substituting $(38)$ in $(36)$ and using $(31)$ and $(39)$ we get the
radial wave equation,
\begin{equation}\label{}
 \frac{d^2R}{dr^2}+\frac{f'(r)}{f(r)}\frac{dR}{dr}+\left[
\frac{\omega^2}{f(r)^2}-\frac{(\frac{2M}{r^3}+
   \frac{\gamma}{r}+\frac{2\Lambda}{3}+\frac{l(l+1)}{r^2})}{f(r)}\right]R=0.
\end{equation}
\\ \\ By  using tortoise  coordinate $ x= \int\frac{dr}{f(r)}$,  the
 above  equation  can  be  brought  into the  standard  form\cite{45},
\begin{equation}
\frac{d^{2}R}{dx^{2}}  +  [\omega^{2} -V(r)]R  =0,
\end{equation}
where,
\begin{equation}
V(r)  =  f(r)(\frac{l(l+1)}{r^{2}}  +\frac{f'(r)}{r}).
\end{equation}
The  SdS   black  hole  has  three  singularities  given   by the
roots  of  $f(r) = 0$,  which  are  the  event  horizon, $r_{1}$,
the cosmological horizon, $r_{2}$   and  at  $r_{3}$  =  -($r_{1}$ +
$r_{2})$. The   QNMs   are  defined  as  solutions  of  the  above
equation  with  boundary   conditions:  $R(x) \rightarrow e^{i\omega
x}$  as  $x \rightarrow \infty$   and  $R(x) \rightarrow e^{-i\omega
x}$  as  $x \rightarrow - \infty$ for an $e^{-i\omega t}$ time
dependence that corresponds to ingoing waves at the horizon and out
going waves at infinity. The surface gravity $\kappa_{i}$ at these
singular points are defined as,
\begin{equation}
\kappa_i =\frac{1}{2}\frac{\partial f}{\partial r}|_{r\rightarrow
    r_i}.
\end{equation}

In  the  present  study  we  are  using  improved AIM   for  finding
the  QNMs  of  the dRGT black  hole  and  hence  it  is convenient
to make a change  of  variable  as  $\xi=1/r$ in $(40)$  leading to,

\begin{equation}
   \frac{d^2R}{d\xi^2}+\frac{p'}{p}\frac{dR}{d\xi}+\left[
\frac{\omega^2}{p^2}-\frac{l(l+1)+(2M\xi+
   \gamma/\xi+\frac{2\Lambda}{3\xi^2})}{p}\right]R=0,
\end{equation}
where,
\begin{eqnarray}
  p &=& -2M\xi^3+\xi^2(1+\zeta)+\gamma\xi+\frac{\Lambda}{3}, \\
  p' &=& -6M\xi^2+\gamma+2\xi(1+\zeta).
\end{eqnarray}
In de Sitter space, the radial equation has got 3 singularities and
these are represented as $\xi_1$ (Event horizon), $\xi_2$
(Cosmological horizon) and $\xi_3 =-\left(
\frac{\xi_1\xi_2}{\xi_1+\xi_2}\right)$ and hence we can write
\cite{ref11,15},
\begin{equation}
    e^{i\omega \xi}=(\xi-\xi_1)^{\frac{i\omega}{2\kappa_1}}%
    (\xi-\xi
    _2)^{\frac{i\omega}{2\kappa_2}}(\xi-\xi_3)^{\frac{i\omega}{2\kappa_3}},
\end{equation}
The idea is to scale out the divergent behavior at the cosmological
horizon first and then rescale at the event horizon for a convergent
solution. Now to scale out the divergent behavior at cosmological
horizon, we take,
\begin{equation}\label{}
    R(\xi)=e^{i\omega \xi}u(\xi).
\end{equation}
The master equation given by $(44)$ then takes the form,
\begin{equation}\label{}
    p u''+ (p'-2 i \omega)u'-\left[l(l+1)+\left(2
    M\xi+\gamma/\xi+\frac{2\Lambda}{3\xi^2}\right)\right]u=0.
\end{equation}
The correct scaling condition of QNM at the event horizon implies,
\begin{equation}\label{}
    u(\xi)=(\xi-\xi_1)^{-\frac{i\omega}{2\kappa_1}}\chi(\xi).
\end{equation}
The master equation then can be viewed of the form as,
\begin{equation}\label{}
    \chi''=\lambda_0(\xi)\chi' +s_0(\xi)\chi,
\end{equation}
where $\lambda_0$ and $s_0$ are the coefficients of the second order
differential equation. It can be seen from $(49)$ that the
coefficient of $u'$ includes the frequency $\omega$. Therefore the
quantization condition given by $(13)$ can be used to find out the
$\omega$ of $(49)$ by iterating to some $n$ maximum. For calculating
the QNMs, we have used the MATHEMATICA NOTEBOOK given in the
reference \cite{46}. Initially the QNMs are calculated for the SdS
by making $\gamma=\zeta=0$ and the results are compared with
reference\cite{ref1,ref2} in Table $1$. It can be seen that the
results agree quite well with those found in the existing
literature.
\begin{table}[h]
\caption{Column 2 shows QNMs calculated for $\gamma=\zeta=0$ for
different values of $\Lambda$ shown in column 1. These are compared
with the SdS case calculated in \cite{ref1} shown in column 4. The
results are found to agree quite well. }
\begin{tabular}{|l|c|c|c|}
  \hline\hline
  $\Lambda$(for dRGT) & $\omega_{AIM}$ &$\Lambda$(for SdS)& $\omega_{WKB}$ \\
  \hline\hline
  0     & 0.483644 -- 0.0967588 i & 0     & 0.48364 - 0.09677 i\\
  -0.02 & 0.434585 -- 0.0885944 i & 0.02  & 0.43461 - 0.08858 i \\
  -0.04 & 0.380784 -- 0.0787610 i & 0.04  & 0.38078 - 0.07876 i\\
  -0.06 & 0.320021 -- 0.0668449 i & 0.06  & 0.32002 - 0.06685 i\\
  -0.08 & 0.247470 -- 0.0519043 i & 0.08  & 0.24747 - 0.05197 i\\
  -0.09 & 0.202960 -- 0.0425584 i & 0.09  & 0.20296 - 0.04256 i\\
  -0.10 & 0.146610 -- 0.0306869 i & 0.10  & 0.14661 - 0.03069 i \\
  -0.11 & 0.0461689 -- 0.0063134 i & 0.11   & 0.04617 - 0.00963 i\\
  \hline\hline
\end{tabular}
\end{table}
We have executed $50$ iterations while calculating the QNMs. We have
taken $(1+\alpha+\beta)<0$ while calculating the QNMs so that the
results of the calculations will correspond to that in de Sitter
space.\\ \\
Table $2$ shows the quasi normal frequencies obtained through
improved AIM method. The values of $\alpha$ and $\beta$ are chosen
so that $\Lambda$ remains negative. We have chosen the values
$M=c=1$ in these calculations. The table shows the quasinormal modes
calculated for $m_g=0.8$ and $m_g=1$ respectively for the same range
of $\alpha$ and $\beta$ values. It can be seen that for the same
$\alpha$ and $\beta$, increasing the value of $m_g$ increases the
magnitude of the cosmological constant, which is obvious from
$(32)$. Also as $m_g$ increases, the quasinormal frequencies are
seen to be increasing in magnitude for both $l=2$ and $l=3$ modes.
As for every $m_g$, both the real and imaginary parts of the
quasinormal frequencies are seen to be continuously increasing in
magnitude as $\Lambda$ increases. Comparing these quasinormal
frequencies with Table $1$, it can be seen that the values of the
quasinormal frequencies when $m_g$ takes a finite value are higher
in magnitude than when $m_g=0$ which corresponds to a Schwarzschild
case.
\begin{table}
\caption{\emph{Quasinormal modes of black hole for massless scalar
perturbations calculated by AIM (with $50$ iterations ) for a class
of de- Sitter dRGT massive gravity for $l=2$ and $l=3$ modes.The
$\alpha$ and $\beta$ values are kept same while QNMs are calculated
by varying the $m_g$ values}}
\begin{tabular}{|ccccc|}
\hline
$m_g=0.8$ &    \hspace{6.3cm}         &            &               & \\
\hline
\end{tabular}
\begin{tabular}{|ccc|c|c|}
\hline \hline
$\Lambda$ & $\gamma$ & $\zeta$ & $\omega (l=2)$ & $\omega (l=3)$  \\
\hline\hline
-0.080 &        -0.80 &         1.9840 &      1.15155 -- 0.348046 i &   1.62914 -- 0.341517 i\\
-0.088 &        -0.80 &         1.9904 &      1.15615 -- 0.350418 i &   1.63572 -- 0.343749 i\\
-0.096 &        -0.80 &         1.9968 &      1.16081 -- 0.352759 i &   1.64237 -- 0.346001 i\\
-0.104 &        -0.80 &         2.0032 &      1.16552 -- 0.355121 i &   1.64910 -- 0.348271 i\\
-0.112 &        -0.80 &         2.0096 &      1.17030 -- 0.357501 i &   1.65590 -- 0.350560 i\\
-0.120 &        -0.80 &         2.0160 &      1.17512 -- 0.359902 i &   1.66278 -- 0.352868 i\\
-0.128 &        -0.80 &         2.0224 &      1.18001 -- 0.362322 i &   1.66974 -- 0.355195 i\\
\hline\hline
\end{tabular}\\
\begin{tabular}{|ccccc|}
$m_g=1.0$ &    \hspace{6.3cm}         &            &               & \\
\hline
\end{tabular}\\
\begin{tabular}{|ccc|c|c|}
\hline \hline
$\Lambda$ & $\gamma$ & $\zeta$ & $\omega (l=2)$ & $\omega (l=3)$  \\
\hline\hline
-0.100 &        -1.00 &         3.1000 &      2.81587 -- 1.049800 i &   3.90051 -- 1.026860 i\\
-0.110 &        -1.00 &         3.1100 &      2.83013 -- 1.057140 i &   3.91984 -- 1.033950 i\\
-0.120 &        -1.00 &         3.1200 &      2.84445 -- 1.064510 i &   3.93924 -- 1.041070 i\\
-0.130 &        -1.00 &         3.1300 &      2.85881 -- 1.071910 i &   3.95870 -- 1.048210 i\\
-0.140 &        -1.60 &         3.1400 &      2.87322 -- 1.079340 i &   3.97823 -- 1.055380 i\\
-0.150 &        -1.75 &         3.1500 &      2.88768 -- 1.086800 i &   3.99781 -- 1.062580 i\\
-0.160 &        -1.90 &         3.1600 &      2.90220 -- 1.094280 i &   4.10746 -- 1.069800 i\\
\hline\hline
\end{tabular}
\end{table}
\subsection{Charged dRGT black hole}
Consider a charged black hole from the class of dRGT massive gravity
with the metric,
\begin{equation}\label{4}
    ds^2=-f(r)dt^2+\frac{dr^2}{f(r)}+r^2d\Omega^{2},\\
\end{equation}
where\cite{39},
\begin{equation}\label{7}
    f(r)=1-\frac{2 M}{r}+\frac{Q^2}{r^2}+\frac{\Lambda}{3}r^2+\gamma
    r+\zeta,
\end{equation}
where $Q$ corresponds to the charge. Proceeding as in section $3.1$,
the wave equation is found as,
\begin{equation}
   \frac{d^2R}{d\xi^2}+\frac{p'}{p}\frac{dR}{d\xi}+\left[
\frac{\omega^2}{p^2}-\frac{(2M\xi-2 Q^2 \xi^{4}+
   \gamma/\xi+\frac{2\Lambda}{3\xi^2})}{p}\right]R=0,
\end{equation}
where,
\begin{eqnarray}
  p &=& Q^2 \xi^{4}-2M\xi^3+\xi^2(1+\zeta)+\gamma\xi+\frac{\Lambda}{3}, \\
  p' &=& 4Q^2 \xi^3-6M\xi^2+\gamma+2\xi(1+\zeta).
\end{eqnarray}
Scaling out the divergent behavior at the event horizon leads to the
master equation,
\begin{equation}\label{}
    p u''+ (p'-2 i \omega)u'-\left[l(l+1)+\left(2
    M\xi-2Q^2\xi^2+\gamma/\xi+\frac{2\Lambda}{3\xi^2}\right)\right]u=0.
\end{equation}
Again, the correct scaling condition of QNMs at the event horizon
implies,
\begin{equation}\label{}
    u(\xi)=(\xi-\xi_1)^{-\frac{i\omega}{2\kappa_1}}\chi(\xi),
\end{equation}
where,
\begin{equation}\label{}
    \begin{split}
    \kappa_1&=\frac{1}{2}\frac{\partial f}{\partial r}|_{r\rightarrow
    r_1},\\
    &=M\xi^2-Q^2\xi^3+\frac{\Lambda}{3}\xi+\frac{\gamma}{2}.
    \end{split}
\end{equation}
The master equation is now in the form of $(1)$ so that the
quantization condition given by $(13)$ can be employed to find out
the QNMs. \\ \\
Table $3$ shows the quasinormal modes calculated using the improved
AIM method for different values of $\alpha$ and $\beta$. We have
chosen the values $M=c=1$ and $Q=0.5$ in these calculations. The
QNMs are studied as in the prevoius section by varying the $m_g$
value while keeping the values of $\alpha$ and $\beta$ the same. It
can be seen that as $m_g$ increases, the real part of the quasi
normal frequency deceases while the magnitude of the imaginary part
increases. For each $m_g$ the quasi normal frequency vary
continuously. A black hole is stable only when the imaginary part in
its Quasi normal spectrum is negative\cite{47}. It is noted while
calculating the Quasinormal modes that the roots of the frequency,
$\omega$ give positive as well as negative imaginary frequencies.
Here we are interested in the stable modes and therefore considered
only the negative imaginary parts of $\omega$. $50$ iterations have
been done for calculating the QNMs.
\begin{table}[h]
\caption{\emph{The Quasinormal modes (after $50$ iterations) for
massless scalar perturbations of a charged black hole for the charge
$Q=0.5$ for the $l=2$ and $l=3$ modes.The $\alpha$ and $\beta$
values are kept same while QNMs are calculated by varying the $m_g$
values}}
\begin{tabular}{|ccccc|}
\hline
$m=0.8$ &    \hspace{6.5cm}         &            &               & \\
\hline
\end{tabular}
\begin{tabular}{|ccc|c|c|}
\hline \hline
$\Lambda$ & $\gamma$ & $\zeta$ & $\omega (l=2)$ & $\omega (l=3)$  \\
\hline\hline
-0.080 &        -0.80 &         1.9840 &      2.43544 -- 0.523799 i &   1.67618 -- 0.168257 i \\
-0.088 &        -0.80 &         1.9904 &      2.43455 -- 0.535763 i &   1.67635 -- 0.180489 i \\
-0.096 &        -0.80 &         1.9968 &      2.43252 -- 0.547233 i &   1.67351 -- 0.195613 i \\
-0.104 &        -0.80 &         2.0032 &      2.42939 -- 0.558215 i &   1.67069 -- 0.209057 i \\
-0.112 &        -0.80 &         2.0096 &      2.42523 -- 0.568718 i &   1.66693 -- 0.222338 i \\
-0.120 &        -0.80 &         2.0160 &      2.42021 -- 0.578624 i &   1.66230 -- 0.235427 i \\
-0.128 &        -0.80 &         2.0224 &      2.41399 -- 0.588313 i &   1.65677 -- 0.248391 i \\
\hline\hline
\end{tabular}\\
\begin{tabular}{|ccccc|}
$m=1.0$ &    \hspace{6.5cm}         &            &               & \\
\hline
\end{tabular}\\
\begin{tabular}{|ccc|c|c|}
\hline \hline
$\Lambda$ & $\gamma$ & $\zeta$ & $\omega (l=2)$ & $\omega (l=3)$  \\
\hline\hline
-0.10 &        -1.00 &         3.1000 &      0.304084 -- 2.99974 i &   0.9866449 -- 4.93190 i\\
-0.11 &        -1.00 &         3.1100&       0.342169 -- 3.05263 i &   1.0195500 -- 5.01834 i\\
-0.12 &        -1.00 &         3.1200 &      0.378347 -- 3.10442 i &   1.0531600 -- 5.10348 i\\
-0.13 &        -1.00 &         3.1300 &      0.413140 -- 3.15511 i &   1.0872800 -- 5.18734 i\\
-0.14 &        -1.60 &         3.1400 &      0.446882 -- 3.20472 i &   1.1219000 -- 5.26998 i\\
-0.15 &        -1.75 &         3.1500 &      0.479812 -- 3.25326 i &   1.1570200 -- 5.35141 i\\
-0.16 &        -1.90 &         3.1600 &      0.512100 -- 3.30072 i &   1.1926600 -- 5.43168 i\\
\hline\hline
\end{tabular}
\end{table}
\section{P-V Criticality of black  holes}                                      
%
\subsection{Black holes in dRGT massive gravity}                                                                     %
In this section we look into the thermodynamic critical behavior of
black holes described by the metric $(31)$ in the extended phase
space. We intend to check whether the black hole exhibits any phase
transition by showing an inflection point in the $P-V$ indicator
diagram. Here, the cosmological constant, $\Lambda$ is treated as
representing a negative pressure \cite{48} as,
\begin{equation}\label{}
    \Lambda=-8\pi P.\\
\end{equation}
For $\gamma=\zeta=0$ the metric given by $(31)$ would lead to the
case of a de Sitter space provided $\Lambda$ is negative. Keeping
this in mind we take,
\begin{equation}\label{}
    \Lambda=8\pi P,\\
\end{equation}
where $P$ is the pressure. The boundary of the black hole is
described by the black hole horizon, $r_h$ and is determined by the
condition, $f(r)|_{r_h}=0$. From this condition, the mass of the
black hole can be expressed in terms of $r_h$ as,
\begin{equation}\label{}
    M = \frac{1}{6} r_h (3 + 3 r_h \gamma + 3 \zeta + r_h^2
    \Lambda),
\end{equation}
and the black hole mass is considered to be the enthalpy of the
system. The thermodynamic volume, $V$ is given by,\cite{49,50}
\begin{equation}\label{}
    V=\frac{\partial M}{\partial P}.
\end{equation}
Varying $(62)$ partially with respect to the pressure P, we get
\begin{equation}\label{}
    V=\frac{4}{3} \pi r_h^3.
\end{equation}
The temperature of the black  hole, described by the metric in
$(31)$, given by the Hawking temperature can be written as
\cite{51},
\begin{equation}\label{}
    \begin{split}
    T &= \frac{1}{4\pi}f'(r_{h}),\\
        &= \frac{1}{4\pi r_h }\left[\frac{2M}{r_h}+\gamma
        r_h+2\frac{\Lambda}{3}r_{h}^2\right],\\
        \end{split}
\end{equation}

Substituting for $M$ from $(62)$ in the above equation and
rearranging it we get an expression for the cosmological constant,
\begin{equation}\label{}
    \Lambda=\frac{4\pi T-2\gamma}{r_h}-\frac{(1+\zeta)}{r_{h}^2}.\\
\end{equation}
But from $(61)$, the cosmological constant can be related to the
pressure as $\Lambda=8\pi P$. Therefore $(66)$ can be written in
terms of $P$ as,
\begin{equation}\label{}
    \begin{split}
    P &= \frac{\Lambda}{8\pi},\\
      &=\left(\frac{T}{2}-\frac{\gamma}{4\pi}\right)\frac{1}{r_h}-%
      \left(\frac{1}{8\pi}+\frac{\zeta}{8\pi}\right)\frac{1}{r_h^{2}},\\
      \end{split}
\end{equation}
Or,
\begin{equation}\label{}
    P=\frac{w_1}{r_h}+\frac{w_2}{r_{h}^2},\\
\end{equation}
where,
\begin{eqnarray}
  w_1 &=& \left(\frac{T}{2}-\frac{\gamma}{4\pi}\right), \\
  w_2 &=& -\left(\frac{1}{8\pi}+\frac{\zeta}{8\pi}\right).
\end{eqnarray}
From $(69)$, $w_1$ can be treated as a shifted temperature. From
$(64)$, thermodynamic volume $V$ is a monotonic function of the
horizon radius $r_h$. and hence $r_h$ can be considered to be
corresponding to $V$. Therefore, $(68)$ can be treated as an
equation of state describing the black  hole. The critical point is
then determined by the conditions,
\begin{equation}\label{}
\frac{\partial P}{\partial r_h}|_{r_h=r_{hc},T=T_c}=0,
\end{equation}
and \\
\begin{equation}\label{}
    \frac{\partial^2 P}{\partial^2 r_h}|_{r_h=r_{hc},T=T_c}=0.
\end{equation}
Substituting for P from $(68)$ in the above differential equation it
is found that the conditions given by $(71)$ and $(72)$ are not
simultaneously satisfied. The condition,
\begin{equation}\label{}
    \frac{\partial P}{\partial r_h}|_{r_h=r_{hc},T=T_c}=0,
\end{equation}
gives the critical horizon as,
\begin{equation}\label{}
    r_{hc}=-\frac{2w_2}{w_1}.
\end{equation}
Evaluation of $\frac{\partial^2 P}{\partial^2
r_h}|_{r_h=r_{hc},T=T_c}$ gives a non zero value which can imply
either a local maximum or a local minimum depending on whether the
value is greater than or less than zero. The critical pressure is
found out by substituting $(74)$ in $(68)$ which gives,
\begin{equation}\label{}
    P_c=-\frac{w_1^{2}}{4w_2}.
\end{equation}
This critical point corresponds to a physically feasible one if
$P_c$ is positive \cite{52}. From $(70)$ it can be seen that this
happens only if $w_2$ is negative irrespective of the sign of $w_1$.
The relation between shifted temperature, $w_1$, critical pressure,
$P_c$ and horizon radius $r_h$ can be found out from $(74)$ and
$(75)$ as,
\begin{equation}\label{}
    \frac{P_c r_{hc}}{w_{1}}=\frac{1}{2}.
\end{equation}
This ratio is called the \lq Compressibility Ratio\rq. The value of
compressibility ratio for a Van der Waal's gas is $0.375$. Hence,
the black hole system, with the Compressibility Ratio given by
$(76)$, can be thought of as behaving like a near Van der Waal's
system. The $P-r_h$ diagram plotted for different shifted
temperature is shown in Figure $1$. In the first figure, the curves
are plotted for $w_2=1$, the curves are seen to show critical
behavior but it likely does not correspond to a physical one
because, from $(75)$, for the above said values of $w_1$ and $w_2$
the critical pressure $P_c$ turns out to be negative for these
curves. The second figure is plotted for $w_2=-0.5$, they show
inflection point but there is no phase transition.
\begin{figure}[h]
  \includegraphics[scale=0.7]{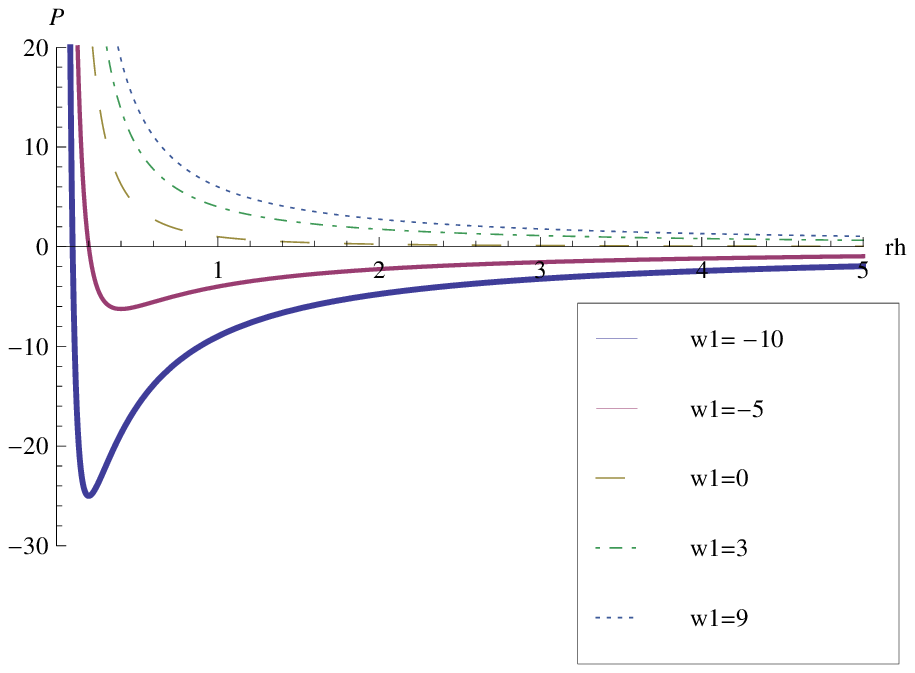}\includegraphics[scale=0.7]{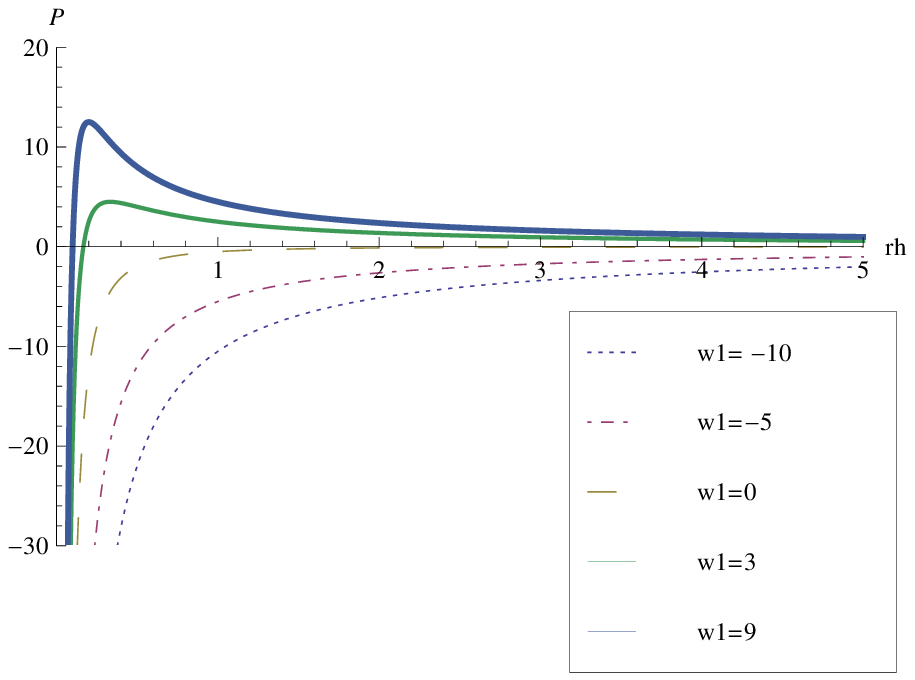}\\
  \caption{In the first figure $p-r_h$ diagram is plotted for the vale $w_2=1$%
  which shows critical behavior and the second figure shows plots for the value %
  $w_2=-0.5$ which shows an inflection.}
\end{figure}
\subsection{Charged dRGT Black  Hole}                                                                  %
Consider a charged black hole with the metric of the form $(56)$.
The Hawking Temperature for this metric can be found out as,
\begin{equation}\label{}
    \begin{split}
    T &= \frac{1}{4\pi}f'(r_{h}),\\
        &= \frac{1}{4\pi r_h }\left[\frac{2M}{r_h}-\frac{2Q^2}{r_h^2}+\gamma
        r_h+2\frac{\Lambda}{3}r_{h}^2\right].\\
        \end{split}
\end{equation}
From the above equation, the equation of state can obtained
proceeding  as described in Section $4.1$. The mass, $M$ of the
black  hole can be written in terms of the horizon radius $r_h$ as,
\begin{equation}\label{}
    M = \frac{1}{6r_h}(3Q^2 +  3r_h^2 (1+\zeta) +3 r_h^3 \gamma +
    r_h^4\Lambda),
\end{equation}
Substituting $(78)$ in $(77)$ we get,
\begin{equation}\label{}
    \Lambda=\frac{4\pi T-2\gamma}{r_h}-\frac{(1+\zeta)}{r_{h}^2}+\frac{Q^2}{r_h^4}.\\
\end{equation}
Writing this equation in terms of $P$,\\
\begin{equation}\label{}
    \begin{split}
    P &= \frac{\Lambda}{8\pi},\\
      &=\left(\frac{T}{2}-\frac{\gamma}{4\pi}\right)\frac{1}{r_h}-%
      \left(\frac{1}{8\pi}+\frac{\zeta}{8\pi}\right)\frac{1}{r_h^{2}}+\frac{Q^2}{8\pi r_h^4},\\
      \end{split}
\end{equation}
Or,
\begin{equation}\label{}
   P=\frac{w_1}{r_h}+\frac{w_2}{r_{h}^2}+\frac{w_3}{r_h^4},\\
\end{equation}
where,\\
\begin{eqnarray}
  w_1 &=& \left(\frac{T}{2}-\frac{\gamma}{4\pi}\right), \\
  w_2 &=& -\left(\frac{1}{8\pi}+\frac{\zeta}{8\pi}\right),\\
  w_3 &=& \frac{Q^2}{8\pi}.
\end{eqnarray}
$(81)$ describes the equation of state. The critical point is then
determined by the conditions,
\begin{equation}\label{}
\frac{\partial P}{\partial r_h}|_{r_h=r_{hc},T=T_c}=0,
\end{equation}
and\\
\begin{equation}\label{}
    \frac{\partial^2 P}{\partial^2 r_h}|_{r_h=r_{hc},T=T_c}=0.
\end{equation}
Unlike in the Section $4.1$, it is found that $(85)$ and $(86)$ are
simultaneously satisfied which gives the solutions, for critical
horizon as,
\begin{equation}\label{}
    r_{hc}=\sqrt{-\frac{6w_4}{w_2}},
\end{equation}
and for the critical temperature as, \\
\begin{equation}\label{}
    w_{1c}=\frac{2}{3}\sqrt{-\frac{2w_2^3}{w_3}}.
\end{equation}
Using $(81)$, $(87)$ and $(88)$, an expression for the critical
pressure can be arrived at as,
\begin{equation}\label{}
    P_c=\frac{w_2^{2}}{12w_3}.
\end{equation}
The relation connecting shifted temperature $w_{1c}$, critical
pressure, $P_c$ and critical horizon radius $r_{hc}$ are found as,
\begin{equation}\label{}
    \frac{P_c r_{hc}}{w_{1c}}=\frac{3}{8},
\end{equation}
which is exactly the same as in the case for a Van der Waal's
system. The P-V diagram  plotted for different shifted temperature
is shown in Figure $2$. In the first figure, the curves are plotted
for $w_2=-10$ and $w_3=1$. The second figure is plotted for $w_2=6$
and $w_3=1$. The first figure shows an inflection point and a phase
transition, but the second does not, as is obvious due to the sign
change of $w_2$.
\begin{figure}[h]
  \includegraphics[scale=0.7]{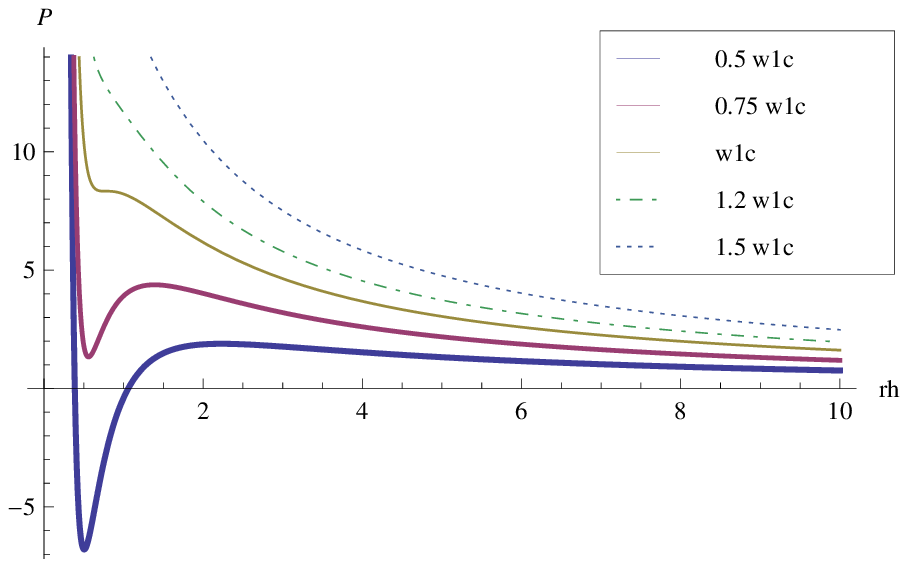}\includegraphics[scale=0.7]{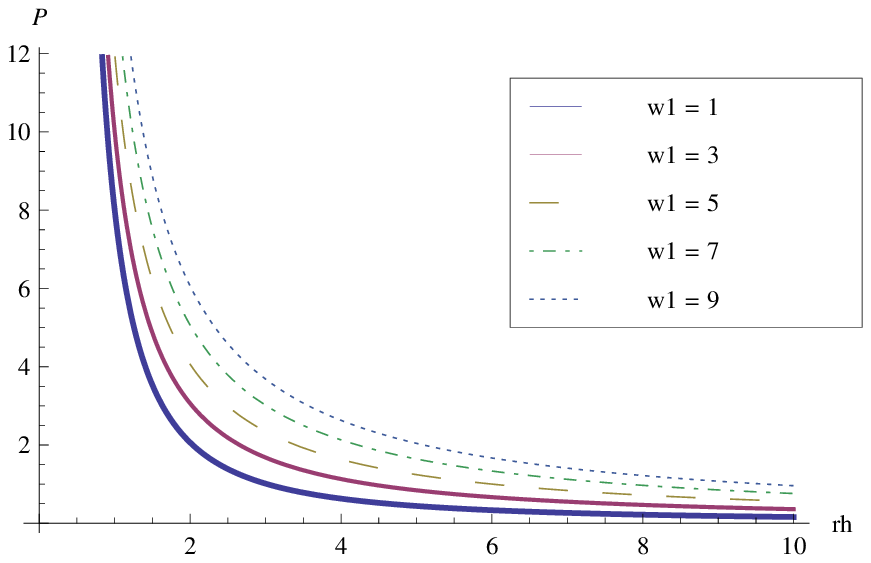}\\
  \caption{In the first figure $p-r_h$ diagram is plotted for the vale $w_2=-10$%
  and $w_3=1$ which shows a phase transition and the second figure shows plots for the value %
  $w_2=6$ and $w_3=1$ which does not show any phase transition.}
\end{figure}
\section{Conclusion}
In this paper, the quasinormal modes coming out of massless scalar
perturbations in black hole space-time in a class of dRGT massive
gravity, is studied. We have used the Improved Asymptotic Iteration
Method (Improved AIM) to find out the QNMs in the de Sitter space.
We have done $50$ iterations for calculating the QNMs. The Quasi
normal modes are studied by varying the massive parameter, $m_g$. It
is found that as $m_g$ increases the magnitude of the quasi normal
frequencies increase for neutral black hole. These QNMs are also
higher in magnitude compared to the SdS case. It is also found that
as $\gamma$ and $\zeta$ tend to zero, the results converge to the
SdS case. For a charged black hole, the real part of the quasi
normal frequency decreases and the magnitude of imaginary part
increases as $m_g$ is increased.  \\ \\
The $P-V$ criticality in the extended phase space of the aforesaid
black holes are also determined. The neutral black holes show a near
Van der Waal behavior with the compressibility ratio of $0.5$. But
it does not show any physically feasible phase transition for the de
Sitter space. The charged black hole on the other hand exactly shows
a Van der Waal's behavior and clearly exhibits a phase transition.
\begin{acknowledgements}
The authors would like to thank the reviewers for their valuable
suggestions. One of us (PP) would like to thank UGC, New Delhi for
financial support through the award of a Junior Research Fellowship
(JRF) during 2010-12 and SRF during 2012-13. VCK would like to
acknowledge Associateship of IUCAA, Pune.
\end{acknowledgements}



\end{document}